\documentclass{article}

\usepackage{arxiv}

\usepackage[utf8]{inputenc} 
\usepackage[T1]{fontenc}    
\usepackage{hyperref}       
\usepackage{url}            
\usepackage{booktabs}       
\usepackage{amsfonts}       
\usepackage{nicefrac}       
\usepackage{microtype}      
\usepackage{lipsum}
\usepackage{graphicx}
\graphicspath{ {./images/} }

\title{Micro frequency hopping spread spectrum modulation and encryption technology}

\author{
 Fanping Du \\
  CMOSTEK Microelectronics Co., Ltd.\\
  \texttt{1511445443@qq.com} \\
  \And
 Pingfang Du \\
  College of Medicine and biological Information Engineering,\\
  Northeastern university\\
  Shenyang, CN 110169 \\
}

\begin{document}

\maketitle
\begin{abstract}
By combining traditional frequency hopping ideas with the concepts of subcarriers and sampling points in OFDM baseband systems, this paper proposes a frequency hopping technology within the baseband called micro frequency hopping. Based on the concept of micro frequency hopping, this paper proposes a micro frequency hopping spread spectrum modulation method based on cyclic frequency shift and cyclic time shift, as well as a micro frequency hopping encryption method based on phase scrambling of baseband signals. Specifically, this paper reveals a linear micro frequency hopping symbol with good auto-correlation and cross-correlation feature in both time domain and frequency domain. Linear micro frequency hopping symbols with different root $R$ have good cross-correlation feature, which can be used in multi-user communication at same time and same frequency. Moreover, there is a linear relationship between the time delay and frequency offset of this linear micro frequency hopping symbol, making it suitable for time delay and frequency offset estimation, also for ranging, and speed measurement. Finally, this paper also verifies the advantages of micro frequency hopping technology through an example of a linear micro frequency hopping spread spectrum multiple access communication system. The author believes that micro frequency hopping technology will be widely used in fields such as the Internet of Things, military communication, satellite communication, satellite positioning, and radar etc.  
\end{abstract}


\section{Micro frequency hopping concept}
\begin{figure} 
    \centering
    \includegraphics{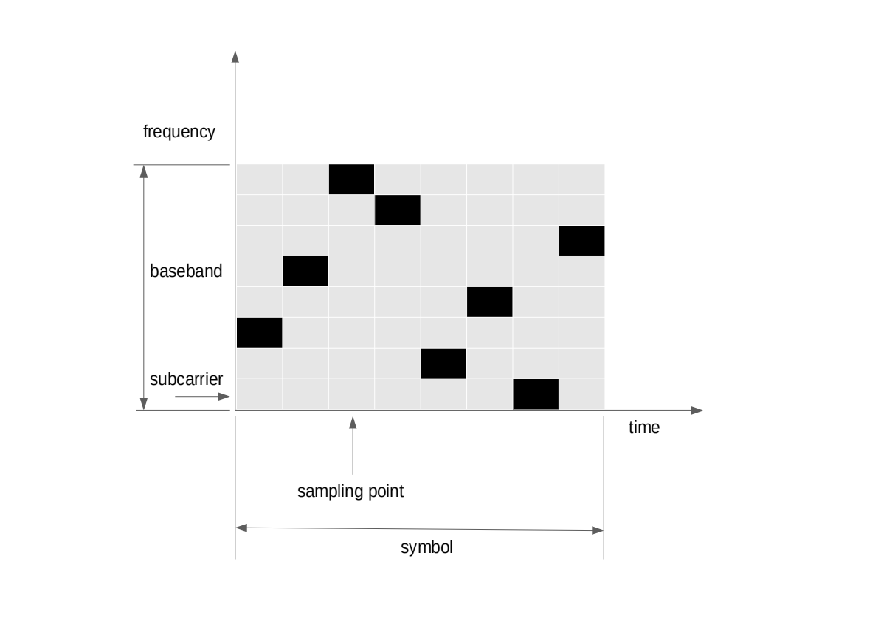}
    \label{fg:mFHP}
    \caption{micro frequency hopping pattern}
\end{figure}

During World War II, the film star "Hedy Lamarr" was inspired by the piano to propose frequency hopping communication technology, which improved the anti-interference ability and confidentiality of communication. On this basis, various frequency hopping spread spectrum technologies such as CDMA, OFDMA, etc. gradually became the mainstream of modern communication technology, such as commonly used 4G, WIFI, etc. On the other hand, in order to more efficiently utilize the baseband spectrum resources, OFDM systems finely divide the baseband time-frequency resources into subcarriers and sampling points. However, OFDM systems use time-frequency resource blocks (RBs) as the basic unit for frequency hopping and do not fully utilize frequency hopping technology. In order to use the spectrum more efficiently, we combine the idea of frequency hopping with the concepts of subcarriers and sampling points in OFDM baseband systems and propose the concept of micro frequency hopping.\\ 
Firstly, micro frequency hopping utilizes the concept of OFDM subcarriers and sampling points to divide the time-frequency resources of the baseband, in the frequency domain the minimum frequency unit is subcarrier frequency point and in the time domain the minimum time unit is sampling point, i.e. micro frequency hopping is a frequency hopping technology that uses the sampling point as the minimum time unit and the subcarrier of the baseband as the minimum frequency unit within the baseband.\\ 
Furthermore, we define a micro frequency hopping pattern, which refers to a time-frequency resource occupancy pattern in the time-frequency resource matrix composed of all subcarriers frequency points within the baseband bandwidth range and all sampling points within a symbol time range, in a micro frequency hopping pattern any sampling point occupies only one subcarrier frequency point.\\
In general, the default baseband bandwidth is equal to the sampling frequency, which means that the number of sampling points for a symbol is the same as the number of subcarrier frequency points in the baseband. Figure (\ref{fg:mFHP}) shows an example of a micro frequency hopping pattern with a symbol length and frequency points of $8$.\\

For example, a micro frequency hopping pattern of size M can be expressed in MATLAB language as follows:\\
\begin{equation} \label{eq:hoppingPattern}
hoppingPattern = randperm(M) - 1
\end{equation}
Here, $hoppingPattern$ is a micro frequency hopping pattern, $M$ is the size of the micro frequency hopping pattern, $randperm$ is a random permute function of M frequency points, according to Matlab rules, the index of the sequence starts from $1$ and needs to be reduced by $1$. Therefore, the micro frequency hopping pattern can be regarded as a frequency point sequence in baseband.\\ 
Furthermore, we define micro frequency hopping symbols as signals generated base on time-frequency points described by micro frequency hopping patterns. Since the accumulation of frequency over time is phase, the phase of the micro frequency hopping symbol at each sampling point can be obtained by accumulating the frequency values of the micro frequency hopping pattern. In other words, the micro frequency hopping symbol can also be regarded as a phase sequence signal with the sampling point as the minimum time unit.\\ 
Expressed in MATLAB language as follows:\\
\begin{equation} \label{eq:hoppingPhase}
hoppingPhase = cumsum(hoppingPattern)/M
\end{equation}
\begin{equation} \label{eq:hoppingSymbol}
hoppingSymbol = exp(2i*pi*HoppingPhase)
\end{equation}
Here, $hoppingPhase$ is the micro frequency hopping phase, $cumsum$ is cumulative sum function, $hoppingSymbol$ is the micro frequency hopping symbol, $exp$ is an exponential function, $i$ is the imaginary unit, and $pi$ is $\pi$. As shown in formula (\ref{eq:hoppingSymbol}), the micro frequency hopping symbol is a complex exponential signal with a constant modulus.\\ 
With the basic concept of micro frequency hopping, we can use micro frequency hopping symbols to implement micro frequency hopping spread spectrum modulation and micro frequency hopping encryption.\\

\section{Micro frequency hopping spread spectrum modulation technology}
\subsection{Micro frequency hopping modulation}
\begin{figure} 
    \centering
    \includegraphics{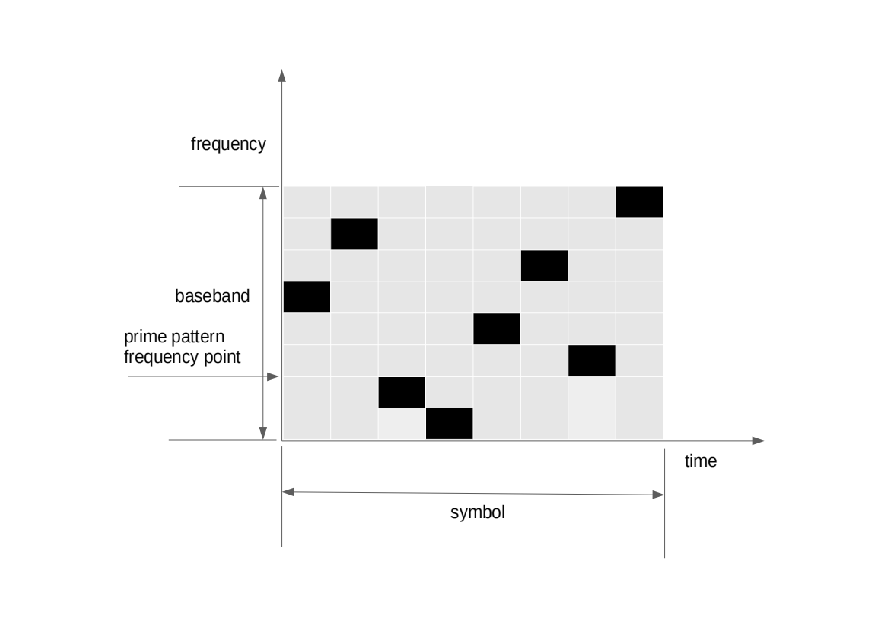}
    \caption{cyclic frequency shift modulation}
    \label{fg:CFSM}
\end{figure}

\begin{figure} 
    \centering
    \includegraphics{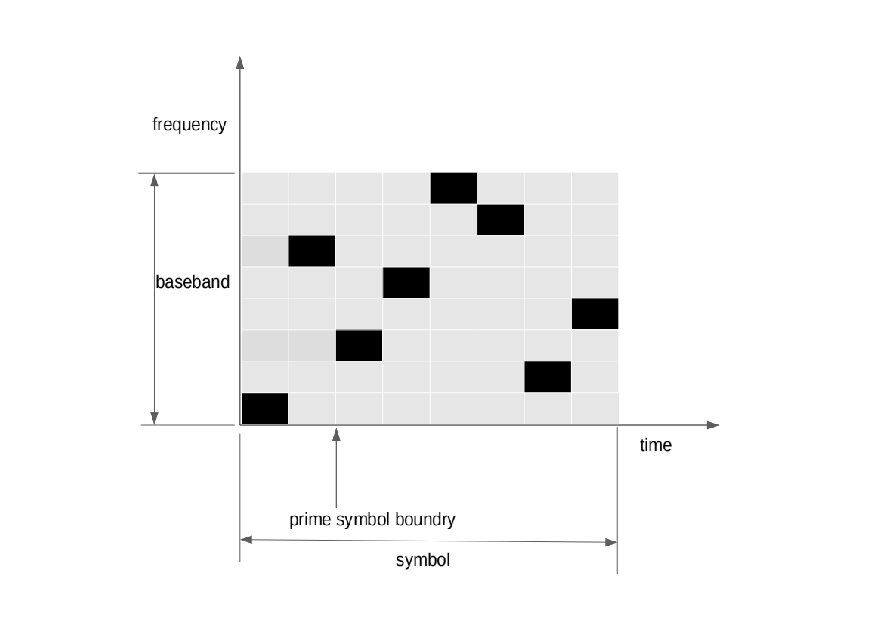}
    \caption{cyclic time shift modulation}
    \label{fg:CTSM}
\end{figure}

Although we can use different micro frequency hopping symbols generated by different micro frequency hopping patterns to representing different information, for simplicity, this paper proposes using cyclic frequency shift or cyclic time shift of the same primary  micro frequency hopping pattern to generate micro frequency hopping symbols representing different information to achieve micro frequency hopping spread spectrum modulation. \\
Figure (\ref{fg:CFSM}) is a schematic diagram of micro frequency hopping cyclic frequency shift spread spectrum modulation, and Figure (\ref{fg:CTSM}) is a schematic diagram of micro frequency hopping cyclic time shift spread spectrum modulation. In this example, the primary  micro frequency hopping pattern is shown in Figure (\ref{fg:mFHP}), and the modulated data represents "$2$", here each micro frequency hopping symbol can modulate $\log_2{8} = 3$ bits of information, and the gain of micro frequency hopping spread spectrum modulation is $10\log_{10}{(8/3)} = 4.26dB$. The cyclic here refers to the frequency shift requiring modulo the baseband bandwidth, and the time shift requiring modulo the symbol time.\\

Taking micro frequency hopping cyclic frequency shift modulation as an example, expressed in MATLAB language as follows:\\ 
\begin{equation} \label{eq:hoppingData}
hoppingData = exp(2i*pi*cumsum(mod(hoppingPattern+data,M)/M)
\end{equation}
Because the period M of a complex exponential function is equivalent to the modulo M, formula (\ref{eq:hoppingData}) can be directly abbreviated as:\\
\begin{equation} \label{eq:hoppingData2}
hoppingData = exp(2i*pi*cumsum(hoppingPattern+data)/M)
\end{equation}
Here, $hoppingData$ is the symbol of micro frequency hopping modulated data, $mod$ is modulo function, and $data$ is the data to be modulated. According to the rules of Matlab, adding a sequence to a constant is equal to adding each data in the sequence to this constant. Cyclic frequency shift refers to taking the modulo of the micro frequency hopping pattern size M after adding the micro frequency hopping frequency point sequence to the data to be modulated.\\

\subsection{Micro frequency hopping demodulation}
According to the method of micro frequency hopping spread spectrum modulation, this paper proposes a micro frequency hopping spread spectrum demodulation method.\\
Taking micro frequency hopping cyclic frequency shift spread spectrum modulation as an example, the demodulation method adopts frequency domain correlation method. Assuming that the frequency offset and time delay have been correctly compensated, the specific process is as follows:\\ 
1: Dot multiply a symbol sized received signal dot by dot with the conjugate of a local reference symbol to obtain a time domain dot multiplication signal, where the local reference symbol corresponds to the primary micro frequency hopping pattern used by the sender;\\ 
2: Transform the time domain dot multiplication signal into a frequency domain correlation signal through Fourier transform;\\
3: Find the maximum amplitude point of frequency domain correlation signals;\\
4: The index of the maximum value in the frequency domain is the demodulated data.\\
Expressed in MATLAB language as follows:\\
\begin{equation} \label{eq:timeDot}
timeDot = rxSig.*conj(hoppingSymbol)
\end{equation}
\begin{equation} \label{eq:freCorr}
freCorr = fft(timeDot)
\end{equation}
\begin{equation} \label{eq:maxIndex}
[~,maxIndex] = max(abs(freCorr))
\end{equation}
\begin{equation} \label{eq:rxData}
demData = maxIndex - 1
\end{equation}
Here, $timeDot$ is the time domain dot multiplication signal, $rxSig$ is the received signal,$.*$ represents the corresponding dot multiplied, $freCorr$ is the frequency domain correlation signal, $fft$ is the Fourier transform function, $maxIndex$ is the maximum index, $max$ is the maximum function, $abs$ is the absolute function, and $demData$ is the demodulated data.\\ 
The essence of this method is frequency domain correlation, which is corresponding the dot multiplication of the signals in the time domain, after transformed into the frequency domain, the correlation peak position in the frequency domain corresponds to the frequency shift of the frequency domain signal. Similarly, for micro frequency hopping cyclic time shift spread spectrum modulation, time domain circular correlation method can be used for demodulation.\\

\subsection{Linear micro frequency hopping and time-frequency estimation}
\begin{figure} 
    \centering
    \includegraphics{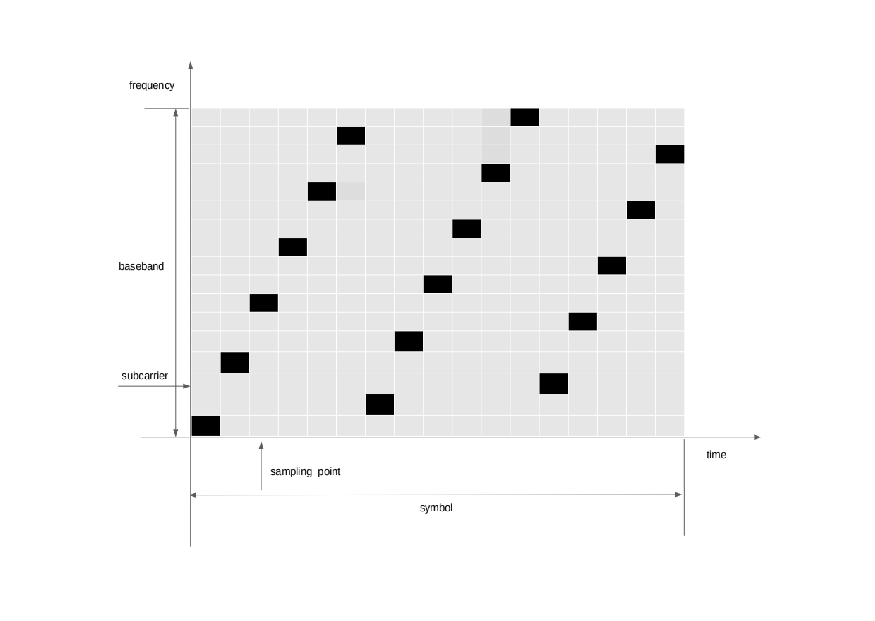}
    \caption{linear micro frequency hopping pattern}
    \label{fg:lmFHP}
\end{figure}

\begin{figure} 
    \centering
    \includegraphics{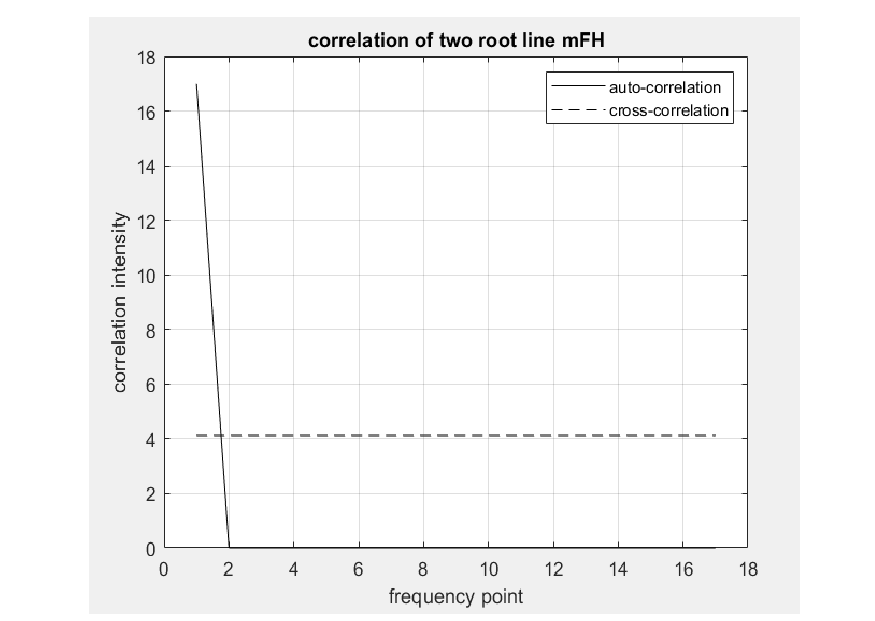}
    \caption{correlation of two roots line micro frequency hopping symbol}
    \label{fg:cor-lmFHS}
\end{figure}

From the micro frequency hopping spread spectrum modulation method, we know that a primary micro frequency hopping symbol and its cyclic frequency shift symbol are correlated in the frequency domain. Similarly, a primary micro frequency hopping symbol and its cyclic time shift symbol are correlated in the time domain. In order to fully utilize channel capacity, we hope to find a set of micro frequency hopping symbols that are uncorrelated in the time domain or frequency domain to achieve micro frequency hopping multiple access spread spectrum communication, that is, multiple users can use multiple uncorrelated micro frequency hopping symbols to achieve communication at same time and same frequency.\\
Fortunately, there does exist a set of uncorrelated micro frequency hopping symbols in both time and frequency domains, namely linear micro frequency hopping symbols of prime size, its feature is:\\
1: The size of the micro frequency hopping pattern is a prime number $P$, which means that both the number of frequency points and the number of sampling points in the micro frequency hopping pattern are prime numbers;\\
2: The frequency point in the micro frequency hopping pattern varies linearly with the sampling point;\\
3: Linear micro frequency hopping symbols with different linear slopes $R$, also known as the root $R$, are uncorrelated in time domain and frequency domain.\\
Figure (\ref{fg:lmFHP}) shows a linear micro frequency hopping pattern with a size of $17$ and a linear slope of $3$.\\
Figure (\ref{fg:cor-lmFHS}) shows the auto-correlation amplitude spectra of linear micro frequency hopping symbol with the same root $R$ and the cross-correlation amplitude spectra of linear micro frequency hopping signals with different root $R$. The auto-correlation amplitude is $P$ and the cross-correlation amplitude is $\sqrt P$. As can be seen from the Figure (\ref{fg:cor-lmFHS}), the linear micro frequency hopping symbols have good auto-correlation and cross-correlation feature.\\

According to the fact that micro frequency hopping symbols are essentially phase sequence signals, a linear micro frequency hopping symbol can be represented as:\\
\begin{equation} \label{eq:hoppingSymbol2}
hoppingSymbol = exp(2i*pi*cumsum(R*(0:P-1))/P)
\end{equation}

\begin{figure} 
    \centering
    \includegraphics{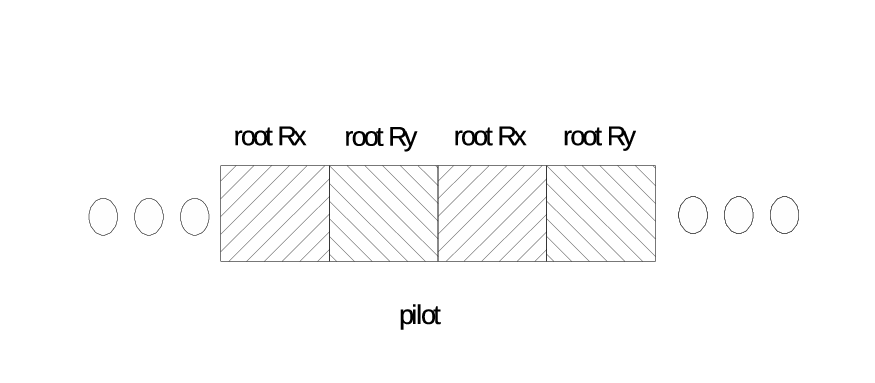}
    \caption{pilot with root $Rx$ and root $Ry$ linear micro frequency hopping symbols}
    \label{fg:RxRy}
\end{figure}

\begin{figure} 
    \centering
    \includegraphics{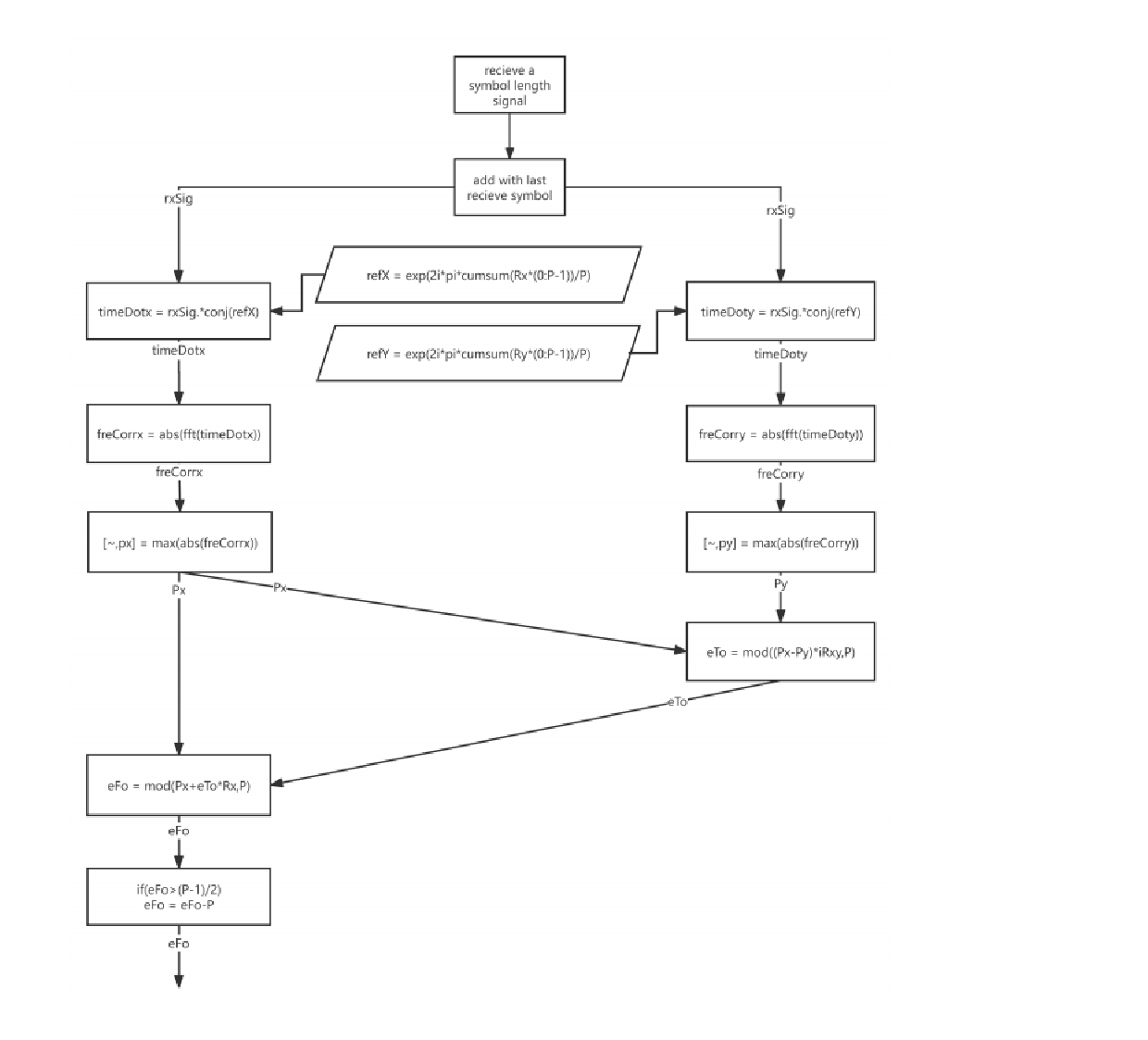}
    \caption{flowchart of time-frequency estimation by linear micro frequency hopping symbols}
    \label{fg:chart}
\end{figure}

Alternatively, according to the sequence summation formula, formula (\ref{eq:hoppingSymbol2}) can be rewritten as:\\
\begin{equation} \label{eq:hoppingSymbol3}
hoppingSymbol = exp(1i*pi*R*(0:P-1).*(1:P)/P)
\end{equation}
Here, $P$ is the size of the micro frequency hopping pattern, which is a prime number, and $R$ is the linear slope or root, with a value range from $1$ to $P-1$.\\
From formula (\ref{eq:hoppingSymbol3}), it can be seen that linear micro frequency hopping symbols are ZC sequence signals, or ZC sequence signals are special linear cases of micro frequency hopping symbols. Therefore, the auto-correlation and cross-correlation feature of linear micro frequency hopping symbols are the same as those of ZC sequence signals.\\
This type of linear micro frequency hopping symbol not only has good time-frequency auto-correlation and cross-correlation feature, but also can be used for time-frequency estimation due to its linear time-frequency relationship. The principle is that the peak position of the time domain auto-correlation of the linear micro frequency hopping symbol is equal to the signal time delay plus $1/R$ times the frequency offset, or the peak position of the frequency domain auto-correlation of the linear micro frequency hopping symbol is equal to the signal frequency offset plus $R$ times the time delay. Here, the time domain auto-correlation is the time domain circular auto-correlation which equivalent to dot multiplied in the frequency domain.\\
Below, this paper proposes a method for time-frequency estimation using linear micro frequency hopping symbols. Since time delay and frequency offset are two independent variables, pilot symbols require at least one pairs of linear micro frequency hopping symbols with different root $R$ to estimate time delay and frequency offset.\\ 
Assuming we choose to arrange the root $Rx$ and $Ry$ of two linear micro frequency hopping symbols in sequence to form pilot symbols, as shown in Figure (\ref{fg:RxRy}).\\
On the receiving side, frequency domain or time domain auto-correlation peak positions can be used for time-frequency estimation. Taking frequency domain auto-correlation as an example, the process is as follows:\\
1: The sender selects two kind linear micro frequency hopping symbols, $Rx$ and $Ry$, and arranges them in sequence to form pilot symbols;\\
2: The receiver sets a symbol cache to add the current received symbol to the previous received symbol;\\
3: Dot Multiply the data of one symbol after addition with the conjugate of two kind local reference symbols in the time domain to obtain a time domain dot multiplication signal,where the two kind local reference symbols are two primary linear micro frequency hopping symbol with the root values $Rx$ and $Ry$;\\
4: Further Fourier transform the time domain dot multiplication signal to the frequency domain to obtain two frequency domain correlation results;
5: Finding the frequency domain correlation peak positions $Px$ and $Py$ of two frequency domain correlation results, and according to that the position of the frequency domain auto-correlation peak position is equal to the frequency offset of the signal plus $R$ times of the time delay, we know:\\
\begin{equation} \label{eq:Px}
Px = Fo - Rx*To
\end{equation}
\begin{equation} \label{eq:Py}
Py = Fo - Ry*To
\end{equation}
Here, $Fo$ is channel frequency offset and $To$ is channel time delay.\\
6: According to formulas (\ref{eq:Px}) and (\ref{eq:Py}), estimate the time delay and frequency offset based on the correlation peak positions in the frequency domain.\\
The estimated time delay value $eTO$ is:\\
\begin{equation} \label{eq:eTo}
eTo = mod((Px-Py)*iRxy, P)
\end{equation}
Here, $mod$ is modulo , and $iRxy$ is the inverse of the difference between the root $Rx$ and root $Ry$ of modulo $P$, i.e. $mod(ixRxy*(Rx-Ry),P)=1$.\\
The estimated frequency offset $eFo$ is:\\
\begin{equation} \label{eq:eFo}
eFo = mod(Px+Rx*eTo, P)
\end{equation}
Or,\\
\begin{equation} \label{eq:eFo2}
eFo = mod(Py+Ry*eTo, P)
\end{equation}
Because frequency offset has positive and negative values, it is necessary to correct the estimated frequency offset value, that is if $eFo>(P-1)/2$, then:\\
\begin{equation} \label{eq:eFo3}
eFo = eFo-P \\
\end{equation}
The flowchart of frequency domain auto-correlation time-frequency estimation is shown in Figure (\ref{fg:chart}).\\
In addition to being used for time-frequency estimation, linear micro frequency hopping can also be used for ranging and speed measurement based on the relationship between distance equal to the speed of light multiplied by time delay and Doppler shift proportional to velocity. Therefore, the linear micro frequency hopping symbol also can be applied in fields such as radar and satellite positioning.\\

\section{Micro frequency hopping encryption}
\begin{figure} 
    \centering
    \includegraphics{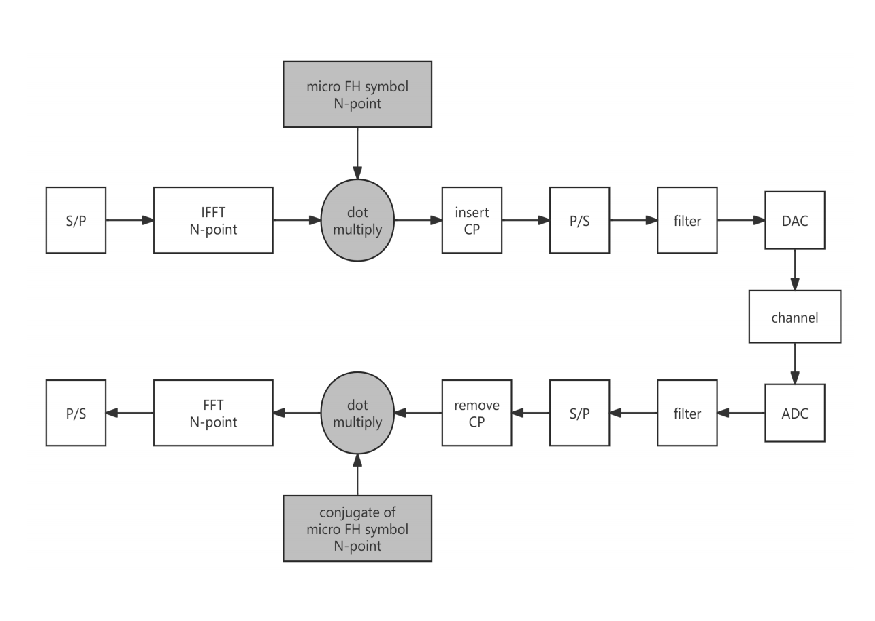}
    \caption{OFDM system with micro frequency hopping phase scrambling}
    \label{fg:OFDM}
\end{figure}

Since the micro frequency hopping symbol can be regarded as a phase sequence signal with sampling points as the minimum time unit, phase scrambling of the signal can also be achieved by multiplying the micro frequency hopping symbol with a general baseband signal.\\
Due to the number of all possible micro frequency hopping patterns with a size of $M$ is $M^M$, and the specific micro frequency hopping patterns used being unknown to third parties, encryption of general baseband signals can be achieved by multiplying the micro frequency hopping symbols with the baseband signal to phase scrambling. Since this encryption encrypts the physical signal itself, rather than the communication data, so micro frequency hopping encryption is a form of physical encryption.\\ 
For example, secure communication can be achieved by multiplying the micro frequency hopping symbol with the baseband signal of a traditional OFDM system to phase scrambling, as shown in Figure (\ref{fg:OFDM}). All subcarriers of the sender are added by IFFT and multiplied with the micro frequency hopping symbol. At this point, the linear phase of each subcarrier signal has been scrambled. The receiver can only recover the initial subcarrier signal by first multiplying the received signal with the conjugate of the same primary micro frequency hopping symbol as the sender, thereby further correctly demodulating the data.\\
Furthermore, this paper provides an example of encrypting using secondary micro frequency hopping on the basis of linear micro frequency hopping symbols.\\
Assuming the primary linear micro frequency hopping symbol is formulas (\ref{eq:hoppingSymbol2}).\\
The secondary micro frequency hopping adopts a random micro frequency hopping pattern as formulas (\ref{eq:hoppingPattern}) .\\
The signal after cyclic frequency shift modulation data with secondary micro frequency hopping is:\\
\begin{equation} \label{eq:hoppingData3}
hoppingData = exp(2i*pi*(cumsum(R*(0:P-1)+data+hoppingPattern))/P)
\end{equation}
Assuming that the received signal noise can be ignored and the frequency offset and time delay have been correctly compensated, if formula (\ref{eq:hoppingSymbol2}) is used as the local reference symbol for frequency domain correlation demodulation, then: 
\begin{equation} \label{eq:timeDot2}
timeDot = hoppingData.*conj(hoppingSymbol) = exp(2i*pi*(cumsum(hoppingPattern+data))/P)
\end{equation}

\begin{figure} 
    \centering
    \includegraphics{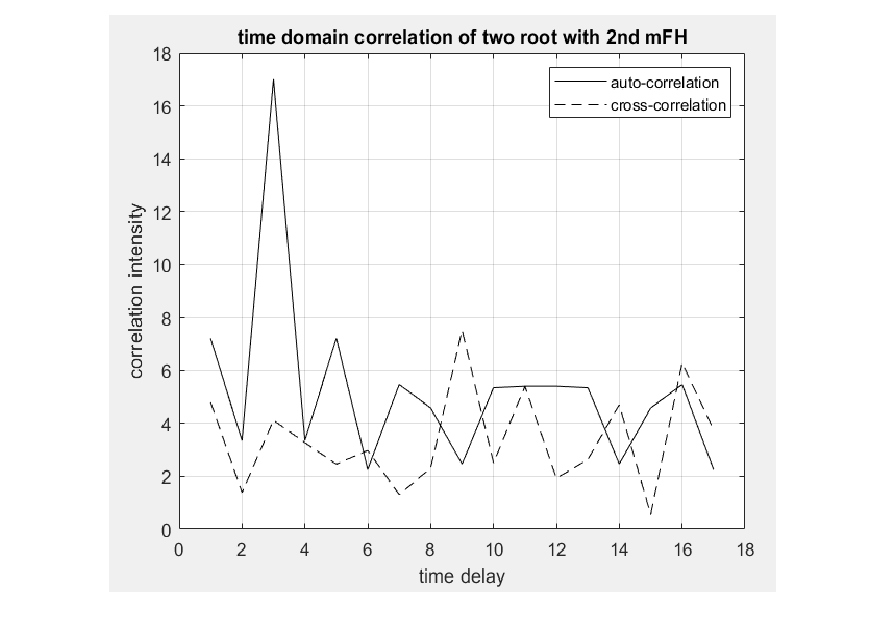}
    \caption{time domain correlation of two roots with secondary micro frequency hopping}
    \label{fg:t-cor-lmFHS-2nd} 
\end{figure}

Formula (\ref{eq:timeDot2}) not only contains single tone 'data' signals, but also a random secondary micro frequency hopping signals, so it is impossible to obtain the correlation peak values by transforming to the frequency domain.\\
If the local reference symbol uses the sum of the linear and random secondary micro frequency hopping patterns, that is, the local reference symbol $sumSymbol$ is:\\
\begin{equation} \label{eq:sumSymbol}
sumSymbol = exp(2i*pi*(cumsum(R*(0:P-1))+hoppingPattern)/P)\\
\end{equation}
After dot multiplying in the time domain:\\
\begin{equation} \label{eq:timeDot3}
timeDot = hoppingData.*conj(sumSymbol) = exp(2i*pi*(cumsum(data))/P)
\end{equation}
The time domain dot multiplication signal is a single tone signal only containing frequency 'data', and after Fourier transform, there exists a frequency domain correlation peak.\\
In addition, the symbol $sumSymbol$ generated by add the random secondary micro frequency hopping pattern to linear micro frequency hopping pattern as formula (\ref{eq:sumSymbol}) has the same frequency domain auto-correlation and cross-correlation feature of linear micro frequency hopping symbol. Unfortunately, in the time domain, although the auto-correlation of $sumSymbol$ has a correlation peak value $P$ , the time delay auto-correlation is no longer zero, and although the cross-correlation does not has a correlation peak value, the cross-correlation value is no longer always equal to $\sqrt P$, as shown in Figure (\ref{fg:t-cor-lmFHS-2nd}).\\

\section{A micro frequency hopping spread spectrum multiple access communication system}

\begin{figure} 
    \centering
    \includegraphics{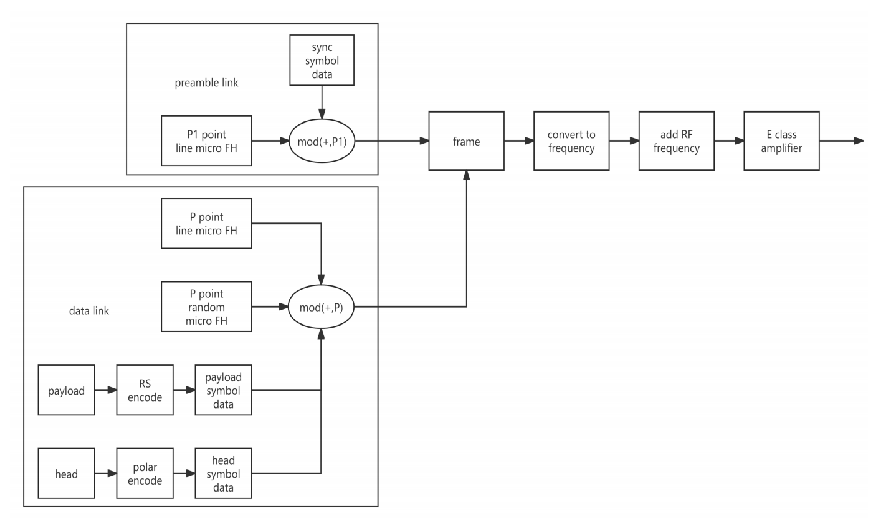}
    \caption{architecture of the micro frequency hopping multiple access communication system}
    \label{fg:MACS}
\end{figure}

Based on the above micro frequency hopping spread spectrum modulation and micro frequency hopping encryption methods, combined with the auto-correlation and cross-correlation of linear micro frequency hopping and its good feature for time-frequency estimation, this paper proposes an example of using linear micro frequency hopping to implement a micro frequency hopping multiple access communication system. \\
The transmission link architecture of the micro frequency hopping multiple access communication system is shown in Figure (\ref{fg:MACS}), which includes a preamble link and a data link. The preamble link includes modules such as linear micro frequency hopping pattern, modulated data, and modulo $P1$ cyclic frequency shift. The data link includes modules such as linear micro frequency hopping pattern, random secondary micro frequency hopping pattern, channel coding, modulated data, and modulo $P$ cyclic frequency shift. The micro frequency hopping multiple access communication system has the following feature:\\
Firstly, the micro frequency hopping multiple access communication system supports multiple users communication at same time and same frequency. Based on the uncorrelated feature of linear micro frequency hopping symbols with different root values, multiple users with different root values can communicate on the same frequency without interference, improving channel utilization. At the same time, there is no need for complex mechanisms such as carrier sense and collision detection between multiple users, and data can be sent asynchronously and randomly, simplifying the design of communication networking protocols.\\
Secondly, the micro frequency hopping multiple access communication system achieves communication security and confidentiality by encrypting the linear micro frequency hopping modulation symbols with secondary micro frequency hopping. The secondary micro frequency hopping pattern is a sequence that can be freely configured and stored by the user. In order to further increase confidentiality, each symbol can read the secondary micro frequency hopping pattern in a specific order. For example, the following formula can be used to read the secondary micro frequency hopping pattern:\\
\begin{equation} \label{eq:xAddr}
xAddr = invE(k*addr,P)
\end{equation}

Here, $xAddd$ is the true read address, $add$ is the sequential address, $invE$ is get inverse element, means $mod(xAddr*(k*addr),P)=1$, it is not a built-in function in Matlab, $k$ is the root of modulo $P$, and each symbol can use different root $k$, so that the secondary micro frequency hopping pattern used for each symbol is in a different order.\\ 
Thirdly, the micro frequency hopping multiple access communication system has higher sensitivity and anti-interference ability. According to the sensitivity calculation formula, the $sensitivity$ of micro frequency hopping modulation is:\\
\begin{equation} \label{eq:sens}
sensitivity = -174 + NF + 10*log10(BW) + EbNo + 10*log10(SF/P)
\end{equation}

\begin{figure} 
    \centering
    \includegraphics{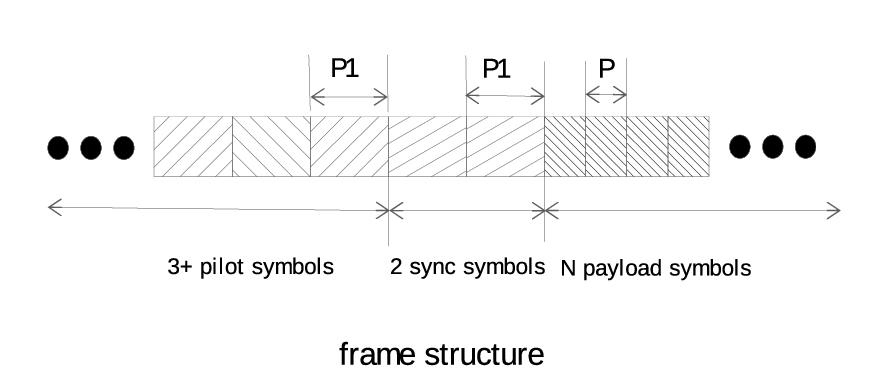}
    \caption{wireless frame structure of the micro frequency hopping multiple access communication system}
    \label{fg:frame} 
\end{figure}

Here, $NF$ is the noise factor, which is about $6dB$, $BW$ is the baseband bandwidth, $P$ is the size of the micro frequency hopping pattern, which is a prime number greater than $2^{SF}$, $SF$ is the spreading factor, which is the number of bits carried by one micro frequency hopping symbol, equal to $\lfloor \log_2{P} \rfloor$, and $EbNo$ is the ratio of energy per bit to noise. Therefore, by adjusting the spreading gain of $10\log_{10}{(P/SF)}$, micro frequency hopping can achieve high sensitivity, especially suitable for long-distance communication, such as LPWAN, satellite communication, etc. Meanwhile, according to the fact that micro frequency hopping patterns are essentially a sequence of frequency points randomly distributed within the baseband bandwidth, it can be inferred that micro frequency hopping symbols have the ability to resist interference from single frequency signals.Furthermore, the data link of the micro frequency hopping multiple access communication system adopts more efficient channel encoding and decoding technologies, where the frame header uses Polar code and the payload uses RS code with variable number of check symbols. Through reasonable channel encoding and decoding, the micro frequency hopping spread spectrum multiple access communication system is only $3dB$ away from the Shannon limit at a frame error rate of $10^{-2}$, further improving sensitivity.\\
Fourthly, the micro frequency hopping multiple access communication system is particularly suitable for low-power communication scenarios. In order to reduce power consumption, the sender can use direct frequency modulation technology, which switches the RF frequency of the transmitted signal in the order of the micro frequency hopping pattern during the sampling time interval of the baseband signal. The RF frequency is the carrier frequency plus the micro frequency hopping frequency point frequency. This modulation technology not only does not require baseband modulation, upsampling, filtering, DUC and DAC circuits, but also can use E-class amplifiers with higher power amplifier efficiency, thus achieving low complexity, low power consumption and high power amplifier efficiency.\\ 
Finally, the micro frequency hopping multiple access communication system achieves more accurate time-frequency estimation by adopting a specially designed wireless frame structure, as shown in Figure (\ref{fg:frame}). The wireless frame includes pilot symbols, synchronization symbols, and data symbols. The pilot symbol and synchronization symbol are generated by the preamble link, with a spreading factor of SF+1. That is, the size $P1$ of the micro frequency hopping symbols used for the pilot symbol and synchronization symbol is a prime number greater than $2^{(SF+1)}$. The pilot symbol is composed of linear micro frequency hopping symbols with roots $R$ and $P1-R$ arranged in sequence. The synchronization symbol is a linear micro frequency hopping symbol with a root $R$ and modulated with synchronization data through cyclic frequency shift. The data symbols are generated by the data link with a spreading factor of SF, which means that the size $P$ of the micro frequency hopping symbols used for the data symbols is a prime number greater than $2^{SF}$. By designing pilot and synchronization symbols with higher spreading gain than data symbols, the micro frequency hopping multiple access communication system can achieve more accurate channel time delay and frequency offset estimation at lower received signals, thereby bringing additional gain to demodulated data symbols.\\
Furthermore, we used SMU200A as the signal source to design the transmission signal of the micro frequency hopping spread spectrum multiple access communication system as shown in Figure (\ref{fg:MACS}) and Figure (\ref{fg:frame}). We also used ADI's ADALM-PLUTO to receive the data, and used Matlab for time-frequency estimation and data demodulation. The verification results showed that the micro frequency hopping spread spectrum multiple access communication system not only has high sensitivity and almost unbreakable confidentiality, but also supports multi-user communication at same time and same frequency, improving channel utilization and simplifying network complexity.\\
This section is just an example of using linear micro frequency hopping symbols and performing cyclic frequency shift modulation. In fact, micro frequency hopping modulation can achieve higher sensitivity or rate through other modulation methods. For example, the micro frequency hopping pattern does not use cyclic frequency shift or cyclic time shift modulation, but instead uses traditional constellation modulation such as DQPSK modulation. One micro frequency hopping symbol only carries $2$ bits, and the spreading gain is $10\log_{10}{(P/2)}$. Therefore, using constellation modulation mode, micro frequency hopping modulation can also achieve higher sensitivity. Furthermore, by superimposing multiple uncorrelated micro frequency hopping symbols on a transmission link, the bit rate of the link can be increased to be suitable for high-speed indoor networking. For example, by superimposing multiple of constellation modulated micro frequency hopping patterns that are mutually cyclic frequency shifted, it is essentially equivalent to an OFDM system superimposing multiple subcarriers and then using the same micro frequency hopping symbol for phase scrambling. The resulting bit rate can be comparable to that of an OFDM system. Therefore, micro frequency hopping communication technology can be flexibly applied to various application scenarios ranging from long-distance to high-speed, from outdoor to indoor.\\

\section{Conclusion}
After leaving the “not only the Fortune Global 500”, I have a chance to enter the field of the Internet of Things. At that time, I was responsible for the development of a chip that may compare with Lora. After analyzing Lora and turbo-FSK technologies, I had a sudden inspiration for micro frequency hopping technology and proposed a series of related patents, this paper is a summary of this series of patents. Although only some of these patents have been implemented in current produced chips, as a technology with independent intellectual property rights, I believe that micro frequency hopping technology will be widely used in the Internet of Things, military communication, satellite communication, satellite positioning, radar and other fields due to its support for multi-user communication at same time and same frequency, and support for signal physical level encryption, strong anti-interference ability, high sensitivity.\\

\bibliographystyle{unsrt}  


\end{document}